\documentstyle[epsf,12pt,aaspp4]{article}

\def\simlt{\mathrel{\hbox to 0pt{\lower 3.5pt\hbox{$\mathchar"218$}\hss}
      \raise 1.5pt\hbox{$\mathchar"13C$}}}
\def\simgt{\mathrel{\hbox to 0pt{\lower 3.5pt\hbox{$\mathchar"218$}\hss}
      \raise 1.5pt\hbox{$\mathchar"13E$}}}

\begin{document}

\title{The Age and Metallicity Range of Early-Type Galaxies in Clusters}
\author{Ignacio Ferreras$^1$, St\'ephane Charlot$^2$,
Joseph Silk$^3$}

\vskip0.5truecm

\affil{
$^1$ Inst. de F\'\i sica de Cantabria, Fac. Ciencias, Av. los Castros s/n,
	39005 Santander, Spain\\
$^2$ Institut d'Astrophysique de Paris, CNRS, 98 bis Boulevard Arago, 75014 Paris, France\\
$^3$ Department of Astronomy, University of California, Berkeley, CA 94720}
\authoremail{
1. ferreras@ifca.unican.es\\
2. charlot@iap.fr\\
3. silk@pac2.berkeley.edu}

\vskip1truecm
\centerline{(Accepted for publication in the {\it Astrophysical Journal})}

\begin{abstract}
We present an unbiased method for evaluating the ranges of ages and
metallicities which are allowed by the photometric properties of the
stellar populations that dominate the light of early-type galaxies
in clusters. The method is based on the analysis of
morphologically-classified early-type galaxies in $17$ clusters at
redshifts $0.3\simlt z\simlt0.9$ and in the nearby Coma cluster using
recent stellar population synthesis models that span a wide range of
metallicities. We confirm that metallicity effects must play a role
in the origin of the slope of the color-magnitude relation for cluster
early-type galaxies. We show, however, that the small scatter of the
color-magnitude relation out to redshifts $z\sim1$ does not formally
imply a common epoch of major star formation for all early-type galaxies.
Instead, it requires that galaxies assembling more recently be on average
more metal-rich than older galaxies of similar luminosity. Regardless of
the true ages and metallicities of early-type galaxies within the allowed
range, their photometric properties and the implied strengths of several
commonly used spectral indices are found to be consistent with {\it
apparently} passive evolution of the stellar populations. Also, the implied
dependence of the mass-to-light ratio on galaxy luminosity is consistent
with the observed trend. The results of our
unbiased analysis define the boundaries in age and metallicity that must
be satisfied by theoretical studies aimed at explaining the formation and
evolution of early-type galaxies in clusters.
\end{abstract}

\keywords{galaxies: evolution --- galaxies: formation --- galaxies: elliptical
--- galaxies: clusters}

\section{INTRODUCTION}

Early-type galaxies in clusters exhibit a linear color-magnitude
(CM) relation indicating that bright galaxies are systematically redder 
than their faint cluster companions (Visvanathan \& Sandage 
1977\markcite{vs77}). This remarkable relation shows very small scatter
($\pm 0.05$ magnitude) in high precision photometry of local clusters such
as Coma and Virgo (Bower, Lucey \& Ellis 1992a\markcite{ble92a}, 1992b,
hereafter BLE92) \markcite{ble92b} and can be extended to clusters at
medium-to-high redshift ($0\simlt z\simlt 1$) (Ellis et al 1997, Stanford,
Eisenhardt \& Dickinson 1998)\markcite{el97}\markcite{sed98}. A first
attempt at explaining the universality of the CM relation involves using
the age of each galaxy as the main determinant of its color. Ageing
stellar populations redden progressively as stars with decreasing initial
mass evolve off the main sequence. Therefore, if the colors of cluster
galaxies are purely controlled by age, the small scatter about the CM
relation implies a nearly synchronous star formation process for all
galaxies of a given mass, while the slope of the CM relation implies
systematically older ages for more massive galaxies. As shown most recently
by Kodama \& Arimoto (1997\markcite{ko97}), such a picture is highly unlikely
because it does not preserve the slope nor the magnitude range of the CM
relation in time.

Another important factor that affects the colors of stellar populations is
metallicity. At fixed age, a more metal-rich stellar population will
appear redder and fainter than a more metal-poor one (e.g., Worthey 1994
\markcite{wo94}). Hence, increasing metallicity at fixed age has a
similar effect on colors as increasing
age at fixed metallicity. This is usually referred to as the {\it
age-metallicity degeneracy} (Worthey 1994\markcite{wo94}).
Several studies have shown that CM relation of cluster elliptical galaxies
could be primarily driven by metallicity effects (Larson 1974\markcite{lar74};
Matteuci \& Tornamb\'e 1987\markcite{mator87}; Arimoto \& Yoshii 1987
\markcite{ari87}; Bressan, Chiosi \& Tantalo 1996\markcite{br96}; Kodama
\& Arimoto 1997\markcite{ko97}). The physical mechanism usually
involved is that of a galactic wind: supernovae-driven winds are expected to be
more efficient in ejecting enriched gas, and hence in preventing more
metal-rich stars from forming, in low-mass galaxies than in massive
galaxies with deeper potential wells. Although age is generally assumed
to be the same for all galaxies in these studies, this has not been
proven to be an essential requirement. In fact, scenarios in which E/S0 galaxies progressively form
by the merging of disk galaxies (Schweizer \& Seitzer
1992\markcite{ss92}) in a universe where structure
is built via hierarchical clustering also predict that the CM relation is
driven primarily by metallicity effects (Kauffmann \& Charlot
1998\markcite{kauf98}). Moreover, age effects could be important
if, for example, there is sufficiently strong feedback from early galaxy
formation to bias the luminous mass distribution of subsequent generations
of galaxies by the heating of intergalactic gas.

In this paper we present a new, more model-independent approach
for evaluating the full range of ages and metallicities allowed by
the spectro-photometric properties of early-type galaxies in clusters.
The method is based on the construction of age-metallicity diagrams
constrained by the colors of early-type galaxies in the nearby Coma
cluster and in 17 clusters observed with the {\it Hubble Space Telescope}
({\it HST}) at redshifts up to $z\approx0.9$ (Stanford et al.
1998\markcite{sed98}). Such an analysis has hitherto been hindered
because of the lack of both accurate stellar libraries for different
metallicities and reliable morphological information on cluster
galaxies at medium-to-high redshifts. Our results can subsequently
be reframed into specific theories of galaxy formation, since they
will be indispensable for any model that seeks to produce galaxies
resembling those actually observed. In \S2 we present the spectral
evolution models used in this paper. The cluster sample is described
in \S3. In \S4 we construct the age-metallicity diagrams allowed by
the observations, and in \S5 we compute the corresponding ranges in
mass-to-light ratio and in several commonly used spectral indices.
We discuss our main conclusions in \S6.

\section{THE MODELS}

We compute the spectral evolution of early-type galaxies using the
latest version of the Bruzual \& Charlot (1998)\markcite{bc98} models
of stellar population synthesis. These span the range of metallicities
$5\times10^{-3} \leq Z/Z_\odot \leq 5$ and include all phases of stellar
evolution, from the zero-age main sequence to supernova explosions for
progenitors more massive than $8\,M_{\odot}$, or to the end of the white
dwarf cooling sequence for less massive progenitors. In addition, the
models predict the strengths of 21 stellar absorption features computed
using the Worthey et al. (1994\markcite{wo94}) analytic fitting functions
for index strength as a function of stellar temperature, gravity and
metallicity. This constitutes the standard ``Lick/IDS'' system that is
often used as a basis for spectral diagnostics in early-type galaxies.
The resulting model spectra computed for stellar populations of various
ages and metallicities have been checked against observed spectra of star
clusters and galaxies (Bruzual et al. 1997\markcite{bru97}; Bruzual \&
Charlot 1998\markcite{bc98}).

The uncertainties in the models are discussed in Charlot, Worthey \&
Bressan (1996).\markcite{cwb96} These can reach up to 0.05~mag in 
rest-frame $B-V$, 0.25~mag in rest-frame $V-K$ and a 25\% dispersion in
the $V$-band mass-to-light ratio. With these uncertainties in mind, we
will concentrate more on understanding the trends seen in the 
observations than on inferring absolute age and metallicity values.
It is worth noting that the most massive elliptical galaxies exhibit
[Mg/Fe] ratios in excess of that found in the most metal-rich stars in the
solar neighborhood (by $\sim0.2-0.3$ dex; see Worthey, Faber, \& Gonzalez
1992\markcite{wo92}). While this may limit the accuracy of the predicted
Mg$_2$ indices of bright elliptical galaxies, the recent models of Bressan
et al. (1998\markcite{br98}) convincingly show that an enhancement in
light elements at fixed total metallicity has virtually no effect on the
other spectrophotometric properties of model stellar populations.

We approximate model early-type galaxies by instantaneous-burst stellar
populations. The reason for this is that we aim at constraining the age
and metallicity ranges of stars dominating the light of early-type galaxies,
whose photometric properties are well represented by instantaneous-burst
populations. In fact, this is true even if the galaxies underwent subsequent
small amounts of star formation or if the epoch of major star formation 
was extended over several billion years (e.g., Fig.~1 of Charlot \& Silk
1994\markcite{cs94}). The predicted colors of our models at fixed age and
metallicity agree well with the results of more refined calculations including
the effects of infall and galactic winds for corresponding values of the mean
age and metallicity (Kodama \& Arimoto 1997, Bressan, Chiosi \& Tantalo
1996).\markcite{ko97} \markcite{br96} For example, adopting metallicities
matching the luminosity--weighted metallicities $\langle \log {Z}/Z_\odot
\rangle$ in Table~2 of Kodama \& Arimoto yields $U-V$ and $V-K$ colors that 
agree to better than 0.05~mag with the results from these authors at an age 
of 15~Gyr.  Such a discrepancy is well within the errors of current population
synthesis models (Charlot et al.  1996\markcite{cwb96}). In the remainder of 
the present paper, the initial mass function (IMF) is taken from Scalo (1986) 
and is truncated at 0.1 and 100$\,M_\odot$.

We use the above  models to compute the locations in the age--metallicity
diagram of stellar populations satisfying  specified spectro-photometric 
properties.  Figure~1 shows four such age--metallicity diagrams 
corresponding to 
imposed values of the $U-V$ and $V-K$ colors and Mg$_b$ and H$\beta$ spectral
indices. These quantities are chosen here because they can be
constrained by many observations of early-type galaxies (\S3 and \S5).
In each panel, the models satisfying the same value of the
spectro-photometric property of interest are related by a continuous line,
different lines corresponding to different imposed values. With this definition,
the slope of a line in the age--metallicity diagrams indicates the
relative sensitivity of the color or index under consideration to age and
metallicity. Vertical lines would correspond to a sensitivity purely to age,
and horizontal lines to a sensitivity purely to metallicity. Figure~1 then 
shows immediately that the $U-V$ and $V-K$ colors and Mg$_b$ index depend
more strongly on metallicity than on age, while the H$\beta$ index depends more
strongly on age than on metallicity. We will return to this point in \S4 and
\S5.

The relative dependence of the spectro-photometric properties of
instantaneous-burst populations on age and metallicity has been previously
investigated by Worthey (1994)\markcite{wo94}. He used the parameter
$\Delta\log t/\Delta\log Z$ at fixed color or index to represent
the ratio of the change $\Delta\log t$ in age needed to counterbalance a 
change $\Delta\log Z$ in metallicity in order to keep that color or index
unchanged. The difference between Worthey's and our approach is that he
computed a single effective value of $\Delta\log t/\Delta\log Z$ for
each spectro-photometric property, while the different lines in Figure~1 
show the behavior of the $\Delta\log t/\Delta\log Z$ slope for
different values of the color or index under consideration. For
comparison, the arrow in each panel of Figure~1 indicates the $\Delta\log
t/ \Delta\log Z$ vector obtained by Worthey (1994)\markcite{wo94}. In each
case, the general agreement with the mean slope of the lines is good.

Table~1 gives a more quantitative comparison between Worthey's (1994)
\markcite{wo94} and our results. We computed linear fits to all lines in
Figure~1 and then a linear fit between the derived slopes and their
corresponding color or index value. The slopes and zero points of these
relations for each spectro-photometric property are listed in columns (2)
and (3) of Table~1. We then evaluated $\Delta\log t/\Delta\log Z$ for four
values of the $U-V$ and $V-K$ colors (and corresponding model predictions
for the Mg$_b$ and H$\beta$ indices) matching the properties of early-type
galaxies at four magnitudes $M_V=-17.5$, $-19.0$, $-20.5$ and $-22.0$ along
the local CM relation (BLE92; see \S3). The agreement with Worthey's (1994)
\markcite{wo94} predictions is seen to be of the order of $\sim 20$ \%. It
is worth noting that our generalized fits deviate significantly from
Worthey's ``3/2 rule'' which takes the $\Delta\log {\rm t}/\Delta\log Z$
ratio to be 1.5 for any color.


 \begin{table*}
   \begin{center}
   \caption{Model comparison}
   \label{table-1}
   \begin{tabular}{c|cc|cccc|c}\hline\hline
  Property & Slope & Zero Point & $M_{\rm V}$= --17.5
  & $M_{\rm V}$= --19.0
  & $M_{\rm V}$= --20.5 
  & $M_{\rm V}$= --22.0 
  & Worthey\\
  \hline
  [$U-V$]        & 0.444 &  0.653 & 1.19  & 1.24  & 1.30  & 1.35  & 1.5\cr
  [$V-K$]        & 0.886 & $-$1.004 & 1.59  & 1.69  & 1.79  & 1.89  & 1.9\cr
  [Mg$_b$]       & 0.118 &  0.976 & 1.33  & 1.36  & 1.40  & 1.45  & 1.7\cr
  [H$\beta$]     & $-$0.092 &  0.645 & 0.47  & 0.48  & 0.49  & 0.50  & 0.6\cr
  \hline\hline
   \end{tabular}
   \end{center}
 \end{table*}

\section{OBSERVATIONAL SUPPORT}

Observational constraints on the photometric properties of early-type
galaxies are taken from the recent sample of Stanford, Eisenhardt \& Dickinson
(1998) \markcite{sed98}. The sample consists of 19 clusters in the redshift
range $0.308\leq z\leq0.895$. These were extracted on the basis of available
{\it HST} imaging from a larger, heterogeneous sample of 46
clusters drawn from a variety of optical, x-ray and radio-selected samples.
The 19 clusters studied by Stanford et al. (1998)\markcite{sed98} were imaged
in the near infrared $J$, $H$ and $K$ passbands. Exposure times in all
passbands were chosen to achieve 5-$\sigma$ detection of objects with the
spectral energy distribution of an unevolved present-day elliptical galaxy
down to 2~mag fainter than $L_\star$ (corresponding to an apparent limiting
magnitude $K=17.6$ and 20.0 at $z=0.308$ and 0.895, respectively). For 17
out of these 19 clusters, photometry is also available in two optical passbands,
referred to as {\it blue} and {\it red}, that were tuned as a function of 
redshift to span the 4000 \AA\  break in the galaxy rest frame spectra. 
This subsample of 17 clusters is of considerable interest to us because it 
sets the most useful constraints on the spectral properties of galaxies. 
Stanford et al.  morphologically classified the galaxies in these clusters 
on the basis of {\it HST} Wide Field and Planetary Camera 2 (WFPC2) images. 
They did not attempt to distinguish between E and S0 galaxies, and for the 
purpose of the present analysis we also consider these galaxy types together 
as a single early-type class.  Table~2 lists the name, redshift, {\it blue} 
and {\it red} passbands and number of E/S0 galaxies for each of the 17 
clusters in the sample.

On the average, cluster membership is expected to be secure for over 85\% of the
E/S0 galaxies in the sample, as estimated either from morphologically-dependent
number counts or from statistical field corrections (see Stanford et al.
1998\markcite{sed98} for a thorough analysis). Furthermore, all clusters
exhibit a tight CM relation with a slope showing no significant change out to
$z\approx0.9$. This has been taken as evidence that all galaxies shared a
common history of star formation (Stanford et al. 1998\markcite{sed98};
Kodama et al. 1998\markcite{ko98}; but see Kauffmann \& Charlot 1998
\markcite{kauf98}). We note that there is an open question
about the membership of some morphologically-selected early-type galaxies with
faint magnitudes and colors far bluer than the CM relation in the clusters
(e.g., Kodama et al. 1998\markcite{ko98}). This can be appreciated most readily
from deep {\it HST} WFPC2 imaging by Ellis et al. (1997) \markcite{el97} of
three of the clusters in Table~2, F1557.19TC, Cl~0016+16 and J1888.16CL.
Although these faint ($22 \simlt I \simlt 23$) galaxies may be field
contaminants, for completeness we should not abandon the possibility that
they are blue cluster early-type galaxies that violate the CM relation. Thus,
they are also included in our analysis.

It is worth pointing out that Stanford et al. (1998)\markcite{sed98} find no
significant difference between the photometric properties of early-type
galaxies in clusters of different richness or x-ray luminosity at a similar
redshift. While this finding does not necessarily imply that all early-type
galaxies were assembled early (Kauffmann \& Charlot 1998\markcite{kauf98}),
it strongly supports analyses like the one presented in \S4 in which
constraints on photometric evolution are drawn from comparisons of the
properties of galaxies in clusters at different redshifts. Also, to avoid 
any bias near the bright end of the CM relation, we have checked that the 
brightest cluster galaxies in the sample of Table~2 match the $K$-band 
absolute luminosity versus redshift relation published by Arag\'on-Salamanca
et al. (1998)\markcite{ar98}.

Finally, to tighten the models at $z\approx0$, we use the CM relation derived
by BLE92 for E/S0 galaxies in the Coma cluster. This takes the forms
$U-V=-0.0819V+2.41$ and $V-K=-0.0743V+4.21$, with standard deviations of
0.055~mag and 0.065~mag in $U-V$ and $V-K$ colors, respectively. To define 
the absolute magnitude scale of the models, the recession velocity of the Coma
cluster is taken to be $cz\approx7186 \,$km~s$^{-1}$ (Han \& Mould
1992\markcite{han92}). We adopt $H_0=60 \,$km~sec$^{-1}$Mpc$^{-1}$ and
$\Omega_0=0.3$, except when otherwise indicated.  The distance modulus to
the Coma cluster is therefore 35.41~mag.

The small scatter of the CM relation at fixed luminosity has been used
by BLE92 to constrain the age range of early-type galaxies in
clusters (see also Ellis et al. 1997\markcite{el97}). They computed the
maximum spread in star-formation epoch allowed by the observed scatter in
$U-V$ color according to the rate of color change $\partial(U-V)/\partial t$
predicted by population synthesis models. The basis of this argument can
be understood from Figure~2. The top panel shows the $U-V$ color evolution
of a stellar population in the time interval $[t,\,t+3\,{\rm Gyr}]$ as a 
function of age $t$ for several metallicities, while the bottom panel
shows the inferred $\partial(U-V)/\partial t$ evolution as a function
of $t$. Since $\partial(U-V)/\partial t$ decreases with model age $t$,
BLE92 found that the older the galaxies, the larger the allowed spread in
star-formation epoch. On the basis of this analysis, BLE92 favored a major
star-formation epoch at $z_F>2$ for early-type galaxies in clusters,
spread over a period of roughly 1~Gyr.

It is crucial to realize that BLE92's argument utilizing
 $\partial(U-V)/\partial t$
is based on the {\it a priori} requirement that at fixed luminosity, all early-type
galaxies in the CM relation have the same metallicity. In fact, the analysis
was conducted using population synthesis models for uniquely solar
metallicity. As the present paper demonstrates, however, at
fixed luminosity the photometric constraints on cluster galaxies allow
a wide range of ages and metallicities (\S4). Hence, the $\partial(U-V)/
\partial t$ argument already includes restrictive hypotheses on the ages and
metallicities of early-type galaxies with respect to the full allowed ranges.
It is not surprising, therefore, that BLE92's conclusions on the epoch of major
star formation for early-type galaxies in clusters are unnecessarily constraining when compared to  
the results of our more complete analysis below.


\begin{table*}
   \begin{center}
   \caption{Cluster Sample}
   \label{table-2}
   \bigskip
   \begin{tabular}{lccccc}\hline\hline
   Cluster & $z_{cl}$ & {\it blue} $-$ {\it red} & N(E/S0)\cr
        & & & \cr
   \hline
   AC 118          & 0.308 & $g-R$ & 38\cr
   AC 103          & 0.311 & $g-R$ & 32\cr
   MS 2137.3-234   & 0.313 & $g-R$ & 21\cr
   Cl 2244-02      & 0.330 & $g-R$ & 24\cr
   Cl 0024+16      & 0.391 & $g-R$ & 39\cr
   GHO 0303+1706   & 0.418 & $g-R$ & 38\cr
   3C 295          & 0.461 & $V-I$ & 25\cr
   F1557.19TC      & 0.510 & $V-I$ & 29\cr
   GHO 1601+4253   & 0.539 & $V-I$ & 42\cr
   MS 0451.6-0306  & 0.539 & $V-I$ & 51\cr
   Cl 0016+16      & 0.545 & $V-I$ & 65\cr
   J1888.16CL      & 0.560 & $V-I$ & 38\cr
   3C 220.1        & 0.620 & $V-I$ & 22\cr
   3C 34           & 0.689 & $V-I$ & 19\cr
   GHO 1322+3027   & 0.751 & $R-i$ & 23\cr
   MS 1054.5-032   & 0.828 & $R-i$ & 71\cr
   GHO 1603+4313   & 0.895 & $R-I$ & 23\cr
   \hline\hline
   \end{tabular}
   \end{center}
\end{table*}

\section{RESULTS}
We now use the models described in \S2 and the observational constraints 
outlined in \S3 to compute the regions allowed in age--metallicity space for
early-type galaxies in clusters. We proceed as follows. We consider each
cluster in Table~2, with redshift $z_{cl}$, and search for all models
which can match the photometric properties of galaxies in that cluster at
$z_{cl}$ and those of Coma galaxies at $z\approx0$. In practice, we
start by considering models that span a full range of metallicities and
with ages at $z=0$ that range between the age of the universe and the
lookback time to redshift $z_{cl}$. Out of these models, we select all
those matching $U-V$ and $V-K$ colors of galaxies in Coma within the scatter
of the observed CM relation (\S3). For a set of $U-V$ and $V-K$ colors this
defines a range of possible $V$ magnitudes. We then compute the predicted
apparent $K$ magnitudes and {\it blue}$-K$, {\it red}$-K$, $J-K$ and $H-K$
colors of the selected models at the cluster redshift $z_{cl}$.

To decide whether a model is acceptable, we compare the predicted photometric
properties at $z_{cl}$ with the observations. For this purpose, the
observed CM relation of each cluster in Table~2 was divided into four apparent
$K$ magnitude bins, roughly 1~mag wide, allowing good sampling from the
brightest to the faintest galaxies (see \S3). For each bin, the standard
deviation around the relation was computed in {\it blue}$-K$, {\it red}$-K$,
$J-K$ and $H-K$
colors. This procedure allows us to account for the increased scatter of the
observed CM relations towards fainter magnitudes. A model is retained if it 
simultaneously falls within $3\sigma$ of the CM relation in all four colors.
The main reason for adopting a $3\sigma$ criterion is that we must allow for
known uncertainties in the spectral evolution models. Uncertainties in the
predicted optical/infrared color evolution over lookback times of a few
billion years can reach a few tenths of a magnitude in current population
synthesis models (Charlot et al. 1996\markcite{cwb96}). This is several times
larger than the typical scatter around the CM relation for the clusters of
Table~2 (see Fig.~5{\it a} of Stanford et al. 1998\markcite{sed98} and Table~1
of Kodama et al. 1998\markcite{ko98}).

Figures~3{\it a}, 3{\it b} and 3{\it c} show the results of our analysis 
for the clusters in the redshift ranges $0.3<z<0.5$, $0.5<z<0.7$ and
$0.7<z<1.0$, respectively. The age--metallicity space is parameterized in
terms of the formation redshift $z_{F}$ and the iron abundance
computed as [Fe/H]$=\log(Z/X)-\log(Z_\odot/X_\odot)$, with $Z_\odot=0.02$
and $Y=2.5Z+0.23$. We separate galaxies into four luminosity bins at
$z\approx0$ by defining apparent magnitude bins in Coma centered on $V=15$,
14.25, 13.5 and 12.75. These correspond roughly to absolute $V$ luminosities
$L_V=L_\ast/4$, $L_\ast/2$, $L_\ast$ and $2L_\ast$, respectively. The
four panels in Figures~3{\it a}--3{\it c} separately show the results for
each of the four luminosity bins. At fixed luminosity, Figure~3 indicates
that the observations of all 17 clusters roughly constrain similar regions
in $(z_F, \,{\rm [Fe/H]})$ space. This is a consequence of the similarity
of the CM relation among all clusters in the sample. An important
discriminant, however, is that the condition $z_{F} >z_{
cl}$ allows more recent formation epochs and higher metallicities for
galaxies in low-redshift clusters than for those in high-redshift clusters.
As expected, different luminosities imply different allowed values of
$z_{F}$ and [Fe/H]. Since faint galaxies on the CM relation are bluer
than bright galaxies, they can be modelled on average by more metal-poor,
and to some extent younger, stellar populations.  We also note that the
metallicity range of model galaxies in Figure~3, $-0.2 \simlt {\rm [Fe/H]}
\simlt +0.2$, compares well with the range observed in nearby E/S0 galaxies
(e.g., Worthey, Faber \& Gonzalez 1992\markcite{wo92}).

The upper panel of Figure~4 shows the areas corresponding to all four
luminosity bins superimposed on a similar diagram for the cluster Cl0016+16.
This cluster, at $z_{cl}=0.545$, has one of the best-defined CM relations
in the sample (65 E/S0 galaxies; see Table~2). Figure~4 readily shows that a 
pure age sequence cannot account for the photometric properties of early-type
galaxies because a single horizontal line cannot cross all four areas 
simultaneously in $(z_F,\,{\rm [Fe/H]})$ space. Alternatively, different 
combinations of age and metallicities can accommodate the data. The simplest 
assumption of a pure metallicity sequence (vertical line) requires that the 
bulk of stars in galaxies formed at redshifts $z_F>2$, otherwise one cannot
account for the photometric properties of the brightest galaxies. 
This confirms earlier conclusions by Ellis et al. (1997\markcite{el97}) based
on similar data for Cl0016+16. As Figure~4 demonstrates, however, there is no 
requirement that the bulk of stars in all early-type galaxies in this cluster 
formed at the same epoch. If the most metal-rich galaxies form most recently,
as might be expected from simple chemical enrichment arguments, then Figure~4 
implies that the dominant stellar populations of all early-type galaxies in 
Cl~0016+16 were already in place at redshifts $z_{F}>2$. On the other hand, 
a wide range of scenarios are also allowed in which faint metal-poor galaxies 
form as recently as $z_{F}\approx1$, i.e., more recently than bright 
metal-rich galaxies. The lower panel of Figure~4 shows that the alternative
cosmology $H_0 =50\,$km~sec$^{-1}$Mpc$^{-1}$ and $\Omega_0=1.0$ would lead on
average to slightly larger $z_{F}$ values for model galaxies because of
the lower age of the universe implied at redshift 1.

We emphasize that the constraints on the formation redshift and metal
abundance of the stellar populations of early-type galaxies derived from
Figure~4 apply to Cl~0016+16 only. As Figure~3{\it a} shows, constraints on
clusters at lower redshifts are consistent with formation redshifts as small
as 0.6 for even the brightest, reddest galaxies. Hence, the apparent
lack of evolution of the slope and scatter of the CM relation in clusters does not
imply by itself a common epoch of star formation for all early-type galaxies.
Instead, what is strictly implied is that new galaxies joining the CM relation
at low redshifts must be more metal-rich than their older cluster companions
of similar luminosity. The required age--metallicity evolution inferred from
Figure~3 corresponds roughly to a 25\% increase in [Fe/H] from $z=1$ to
$z=0.5$, i.e., over a period of 2.5~Gyr for the assumed $H_0=60\,$km~sec$^{
-1}$Mpc$^{-1}$ and $\Omega_0=0.3$. We note that adopting longer timescales of
star formation --- instead of an instantaneous burst --- for early-type galaxies
in clusters would increase the formation redshift of the first stars in model
galaxies with respect to the results of Figure~3. Whatever the adopted history
of star formation, however, the values of $z_{F}$ in Figure~3 are robust limits
on the redshift of the last major event of star formation in early-type galaxies
(see \S2).

Finally, we have explored the possibility that the faint
morphologically-selected early-type galaxies with very blue colors in
Cl~0016+16 were actual cluster members (see \S3). The inferred $z_{F}$
and [Fe/H] ranges for these objects is indicated by the heavy shaded region 
in Figure~4. As expected, the blue colors imply that the galaxies must be
both young and metal-poor. We may speculate that they are young objects which
``violate'' the main sequence CM relation, but which could eventually evolve
towards it as they age and undergo chemical enrichment. However, spectroscopic
confirmation as well as a deeper search for faint blue outliers in clusters
at higher redshifts are needed before we can draw any definitive conclusions
as to their true nature.

\section{APPLICATIONS}

The above analysis has enabled us to determine the areas allowed in $(z_F,
\,{\rm [Fe/H]})$ space for the stellar populations of E/S0 galaxies in
clusters. While we emphasize that real galaxies may occupy only part of
these allowed areas, it is interesting to compute the implied evolution
with redshift of other observable properties of early-type galaxies such
as mass-to-light ratios and spectral indices. Since we do not know which
subareas real galaxies might occupy in Figure~3, in the following we 
investigate properties averaged over the entire allowed areas. 

\subsection{Mass-to-Light Ratios: The Fundamental Plane Revisited}

We first investigate the mass-to-light ratios of the model galaxies
selected in Figure~3. For each cluster, we compute the logarithmic mean
of $M/L_V$ in each luminosity bin at the cluster redshift by averaging over
the $(z_F,\,{\rm [Fe/H]})$ area constrained by the observations. Since
absolute values of $M/L_V$ depend sensitively on the assumed low-mass end
of the IMF, we adopt for all clusters the arbitrary normalization
$\log(M/L_V)=0.8$ at $L_V =L_\ast$ and focus our investigation on the
dependence of the mass-to-light ratio on luminosity. The mean values of
$\log(M/L_V)$ computed in this way are shown as triangles in Figure~5.
Since the mass-to-$V$ light ratio of a stellar population increases at
increasing age and metallicity, brighter galaxies in Figure~5 generally
tend to have larger $\log(M/L_V)$ than fainter galaxies. This is not always
true, however, as can be understood from the dispersion in the results
of Figure~3. 

Also shown in Figure~5 are the results from two recent observational
studies of the fundamental plane for cluster E/S0 galaxies. J{\o}rgensen,
Franx, \& Kj{\ae}rgaard (1996)\markcite{jo96} have parameterized the
mass-to-Gunn~$r$ luminosity ratio, $M/L_r$, in terms of a combination of
half--light radius and central velocity dispersion from observations of
a large sample of 226 E/S0 galaxies in 10 nearby clusters. They conclude
that the half--light radius has a negligible effect on determinations of
$M/L_r$ and show that the data follow the simple relation $M/L_r\propto
\sigma^{0.86}$ with a scatter of only 25\%. The central velocity
dispersion $\sigma$ is tightly correlated with luminosity via the Faber-Jackson
relation (Faber \& Jackson 1976\markcite{fj76}). We use BLE92's calibration
of this relation based on observations of early-type galaxies in the Coma
cluster in order to relate $M_V$ to $\log\sigma$ in Figure~5. We then
adopt the mean $V-r$ colors of galaxies along the CM relation to convert
J{\o}rgensen et al.'s (1996)\markcite{jo96} result into an expression
involving the mass-to-$V$ luminosity ratio. This yields $M/L_V\propto
\sigma^{0.88}$, corresponding to an increase of less than 3\% in
logarithmic slope with respect to the relation derived in Gunn~$r$. The
horizontal shading in Figure~5 then indicates the range of slopes allowed
after accounting for the scatter of J{\o}rgensen et al.'s
(1996)\markcite{jo96} data around the mean relation.

An alternative constraint on the observed range of mass-to-$V$ luminosity
ratios can be obtained from the relation recently derived by Graham \&
Colless (1997)\markcite{gr97} between half--light radius, central velocity
dispersion and mean surface brightness of early-type galaxies. Their 
analysis is based on accurate fitting of the $V$-light profiles of 26 E/S0
galaxies in the Virgo cluster using both homologous ($\propto r^{1/4}$)
and non-homologous ($\propto r^{1/n}$, with $n$ a free parameter) radial
dependence laws. The results indicate a slight but systematic breaking
of the generally assumed homology in the sense that $n$ increases with
increasing half-light radius. We can use the virial theorem and
Faber-Jackson relation to reexpress the results of Graham \& Colless
(1997)\markcite{gr97} in terms of a relation involving $M/L_V$ and $M_V$.
The outcome is shown in the upper and lower panels of Figure~5 for
non-homologous and homologous light profiles, respectively. In each
case, the heavy solid line indicates the mean relation and the slanted
shading represents the allowed range of slopes when the uncertainties 
quoted by Graham \& Colless are included.

Figure~5 shows that the mass-to-light ratios of model galaxies are all
within the range constrained by current observations. The models appear
to be mostly compatible with the Graham \& Colless (1997)\markcite{gr97}
results for non-homologous light profiles. However, we point out that
the model values correspond to purely stellar mass-to-light ratios,
whereas there is some observational evidence for the presence of
dark halos around early-type galaxies. This is inferred via the studies
of HI kinematics (e.g. Franx et al. 1994), x-ray emission (e.g. Forman et
al. 1994\markcite{fo94}), radial velocities of planetary nebulae and
globular clusters (e.g. Mould et al. 1990\markcite{mo90}, Hui et al.
1995\markcite{hui95}), gravitational lensing (e.g. Maoz \& Rix
1993\markcite{mao93}) and measurements of the shape of the stellar
line-of sight velocity distribution (e.g. Carollo et al.
1995\markcite{ca95}). More recently, Rix et al. (1997\markcite{ri97})
have analyzed the velocity profile of the elliptical galaxy NGC~2434
and show that roughly half the mass within an effective radius is
dark. Hence, one must remain cautious in interpreting trends in the
mass-to-light ratios of early-type galaxies on the basis of pure 
population synthesis models.

\subsection{Spectral Indices}

For completeness, we also compute the strengths of several commonly
used spectral indices implied by the above photometric study for 
early-type galaxies in clusters. Most observations in this domain
have concentrated on measurements of H-Balmer, magnesium and 
iron-dominated indices of the Lick system (\S2) such as H$\beta$,
Mg$_2$, Mg$_b$ and MgFe (Worthey et al. 1992\markcite{wo92}; Bender,
Burstein \& Faber 1993\markcite{be93}; Gonzalez 1993\markcite{go93};
Davies, Sadler \& Peletier 1993\markcite{da93}; Ziegler \& Bender
1997\markcite{zi97}). One of the main motivations for studies of
this type is to identify a pair of indices including one mostly 
sensitive to age and one mostly sensitive to metallicity in order 
to break the age--metallicity degeneracy from photometric colors.
The age--sensitive H$\beta$ index, however, can be significantly
contaminated by nuclear emission in E/S0 galaxies, for which 
the corrections are uncertain (e.g., Gonzalez 1993\markcite{go93}).
It is therefore deemed preferable to study higher-order Balmer 
indices such as H$\gamma$ (e.g., Worthey \& Ottaviani
1997\markcite{wo97}).  Also, the apparent systematic increase of
[Mg/Fe] from faint to bright E/S0 galaxies challenges analyses based
on models with solar-scaled abundance ratios (e.g., Worthey et al.
1992\markcite{wo92}; Tantalo, Bressan \& Chiosi 1997\markcite{ta97}
and references therein). Worthey (1995\markcite{wo95}) shows that
the metal-sensitive index C4668 of the Lick system appears to be
significantly less enhanced with respect to iron than Mg in bright
E/S0 galaxies. Thus, C4668 represents a good alternative to
Mg-dominated indices for reliable model investigations of early-type
galaxy spectra (e.g., Kuntschner \& Davies 1998\markcite{ku98}).

By analogy with our approach in \S5.1, we compute mean spectral
indices for each of the four luminosity bins in each cluster
of our sample. The results are shown in Figure~6 for the Mg$_2$,
H$\beta$, Mg$_b$, MgFe, C4668 and H$\gamma_A$ indices of
the Lick system. As expected, there is a general correlation 
between index strength and luminosity, which is more pronounced
for metal-sensitive (Mg$_2$, Mg$_b$, MgFe and C4668) than for
age-sensitive (H$\beta$, H$\gamma_A$) indices. The 
reason for this is that luminous galaxies in Figure~3 are on
average found to be significantly more metal-rich and slightly 
older than faint galaxies. Also, Figure~1 shows that the
sensitivity to age of H-Balmer indices such as H$\beta$ is
significantly weakened after $2-3$~Gyr, when A and B stars have
evolved off the main sequence. For comparison, the solid and
dashed lines in Figure~6 show the locations of passively
evolving model galaxies with $z_{F}=5$ and the metallicities
$Z_\odot$ and $0.4 Z_\odot$, respectively. The mean index strengths
of cluster galaxies appear to evolve roughly along these lines,
supporting the consistency of the stellar populations with
apparently passive evolution. In fact, Ziegler \& Bender
(1997\markcite{zi97}) find that the evolution from $z\approx0$
to $z\approx0.4$ of the correlation existing between the Mg$_b$
index strength and central velocity dispersion of E/S0 galaxies
is consistent with passive evolution. It should be noted that
the model indices computed in Figure~6 are global indices 
averaged over the emission from all stars in a galaxy. Observations
of nearby E/S0 galaxies, however, reveal systematic variations of
the strengths of spectral indices between the central and outer regions
(e.g., Worthey et al. 1992\markcite{wo92}). 

\section{DISCUSSION}

We have shown that the tight photometric constraints on early-type
galaxies in clusters allow relatively wide ranges of ages and
metallicities for the dominant stellar populations. In particular, the
small scatter of the CM relation out to redshifts $z\sim1$ does not
necessarily imply a common epoch of star formation for all early-type
galaxies. It requires, however, that galaxies assembling more recently be
on average more metal-rich than older galaxies of similar luminosity.
In this context it is interesting to mention that, based on the spectral
indices of nearby E/S0 galaxies, Worthey, Trager \& Faber (1996\markcite{wo96})
favor younger ages for more metal-rich galaxies than for metal-poor
ones at fixed velocity dispersion. The results of our unbiased analysis
therefore define the boundaries in age and metallicity that must be satisfied
by theoretical studies aimed at explaining the formation and evolution of
early-type galaxies in clusters.

The constraints obtained here on the age and metallicity ranges of E/S0
galaxies are consistent with conventional models in which the galaxies
all form monolithically in a single giant burst of star formation at high
redshift (e.g., Kodama et al. 1998\markcite{ko98}, and references therein). In
fact, this implies that regardless of the true ages and metallicities of
early-type galaxies within the allowed range, their photometric properties will
always be consistent with {\it apparently} passive evolution of the stellar
populations. As Figure~6 shows, this consistency even extends to  spectral
index strengths.

Our results are also consistent with scenarios in which E/S0
galaxies are formed by the merging of disk galaxies (Schweizer \& Seitzer
1992\markcite{ss92}) in a universe where structure is built through
hierarchical clustering (Kauffmann 1996\markcite{kauf96}; Baugh, Cole \&
Frenk 1996\markcite{ba96}; Kauffmann \& Charlot 1998\markcite{kauf98}). For
such scenarios, Figure~3 constrains the metallicity and epoch of the last major
event of star formation in E/S0 galaxies and their progenitors (see \S2 and
\S4). The ages and metallicities of cluster E/S0 galaxies predicted by 
hierarchical models are found to be consistent with these constraints
(Kauffmann \& Charlot 1998\markcite{kauf98}).

To better assess the origin of E/S0 galaxies in clusters one therefore needs
to appeal to observational constraints other than their spectro-photometric
properties. For example, conventional models of E/S0 galaxy formation are
being challenged by the paucity of red galaxies found at high redshifts in
deep surveys (Kauffmann, Charlot, \& White 1997\markcite{kcw97}; Zepf
1997\markcite{zepf97}). Also, morphological distinction between E and S0
galaxies and the evolution of the morphology-density relation out to moderate
redshifts appear to point to different formation epochs for E and S0 galaxies
(Dressler et al. 1997\markcite{dres97}).

The tightness of the CM relation is proof of a stable process in the
assembly of cluster early-type galaxies. However, as we move towards greater
redshifts, a drastic change is expected at lookback times that approach
the formation of the first E/S0 galaxies. This change can arise as a 
systematic blueing, an increased scatter or a slope flattening in the CM
relation (e.g., Arag\'on-Salamanca et al. 1993; Charlot \& Silk
1994\markcite{cs94}; Kauffmann \& Charlot 1998\markcite{kauf98}). An 
interesting question is raised by the presence of morphologically-selected
early-type galaxies with very blue colors in clusters at moderate redshifts
(\S3 and \S4). If these objects are true cluster members, our analysis shows
that they could be young metal-poor galaxies that will later join the CM
relation. Hence, we need to probe deeper down the galaxy luminosity function
in distant clusters in order to assess whether these objects can have any
fundamental bearing on the origin of early-type galaxies.

\acknowledgments
We thank Adam Stanford for sending us a machine readable list of the cluster
photometry used in this paper. I.F. thanks the CNRS and the Institut
d'Astrophysique de Paris for hospitality and financial support, and also
acknowledges a Ph.D. scholarship from the ``Gobierno de Cantabria.'' J.S.
acknowledges support from NASA, NSF and the Blaise-Pascal chair at the
Institut d'Astrophysique de Paris.



\clearpage
\plotone{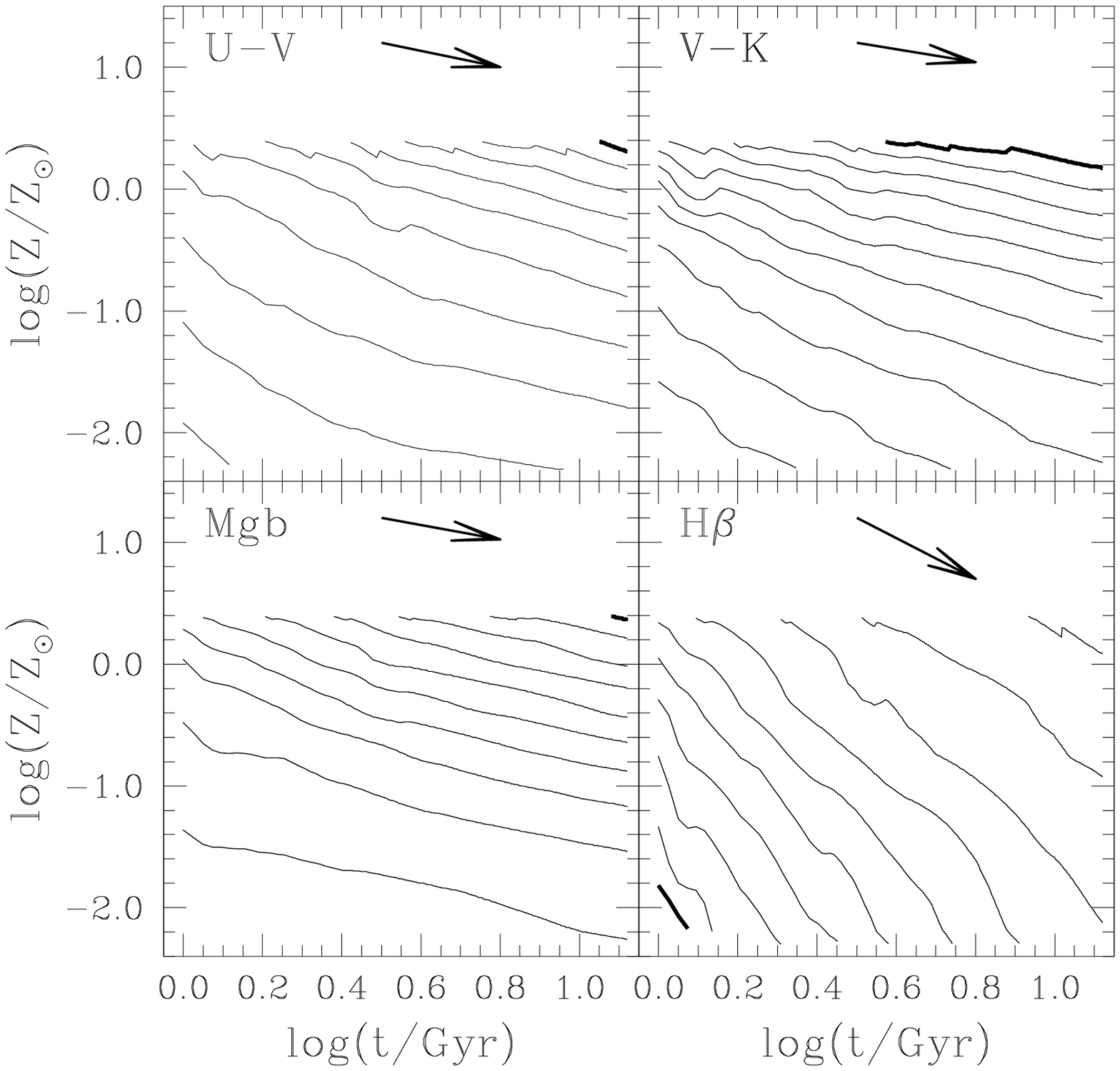}
\vskip-0.5truecm
\figcaption[fig1.eps]{Curves in age--metallicity space corresponding to fixed
values of a specific spectro-photometric index for model early-type galaxies.
{\it Top-left panel}~: $U-V$ color from 0.2 to 2.0, in increments of 0.2.
{\it Top-right panel}~: $V-K$ color from 1.5 to 3.5, in increments of 0.2. 
{\it Bottom-left panel}~: Mg$_b$ index from 0.5 to 5.0, in increments of 0.5.
{\it Bottom-right panel}~: H$\beta$ index from 1.5 to 6.0, in increments of
0.5. In each panel, the curve corresponding to the highest index value is
shown as a thick line. The arrows represent the mean $\Delta\log{\rm age} /
\Delta\log Z$ slopes from Worthey (1994). \label{f1}}

\clearpage
\plotone{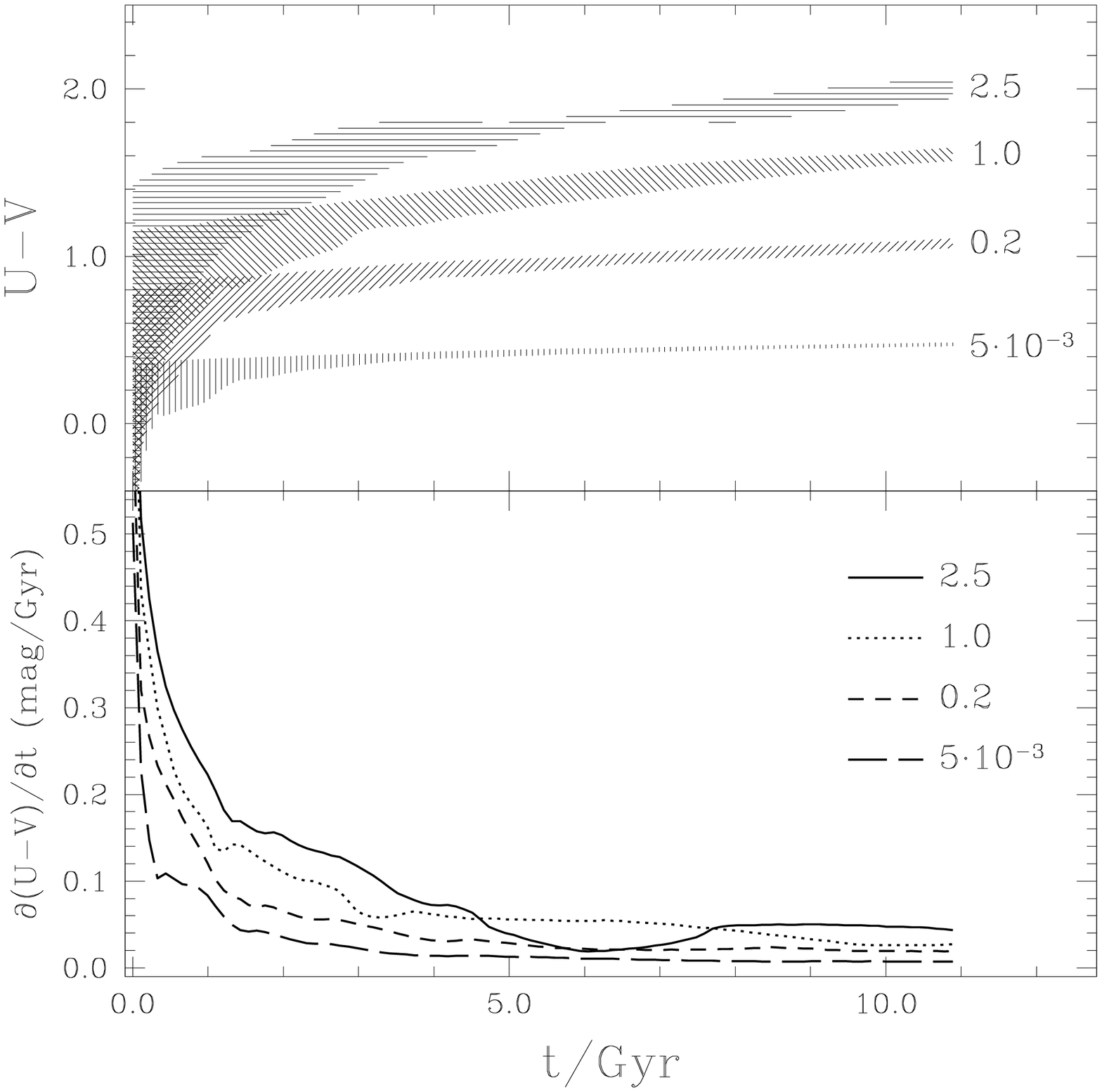}
\figcaption[fig2.eps]
{{\it Top panel}: shaded regions indicate the $U-V$ color evolution of
model early-type galaxies in the time interval $[t,\,t+3\,{\rm Gyr}]$ as
a function of age $t$ for several values of $Z/Z_\odot$, as indicated.
{\it Bottom panel}: corresponding evolution of $\partial(U-V)
/\partial t$. \label{f2}}

\clearpage
\plotone{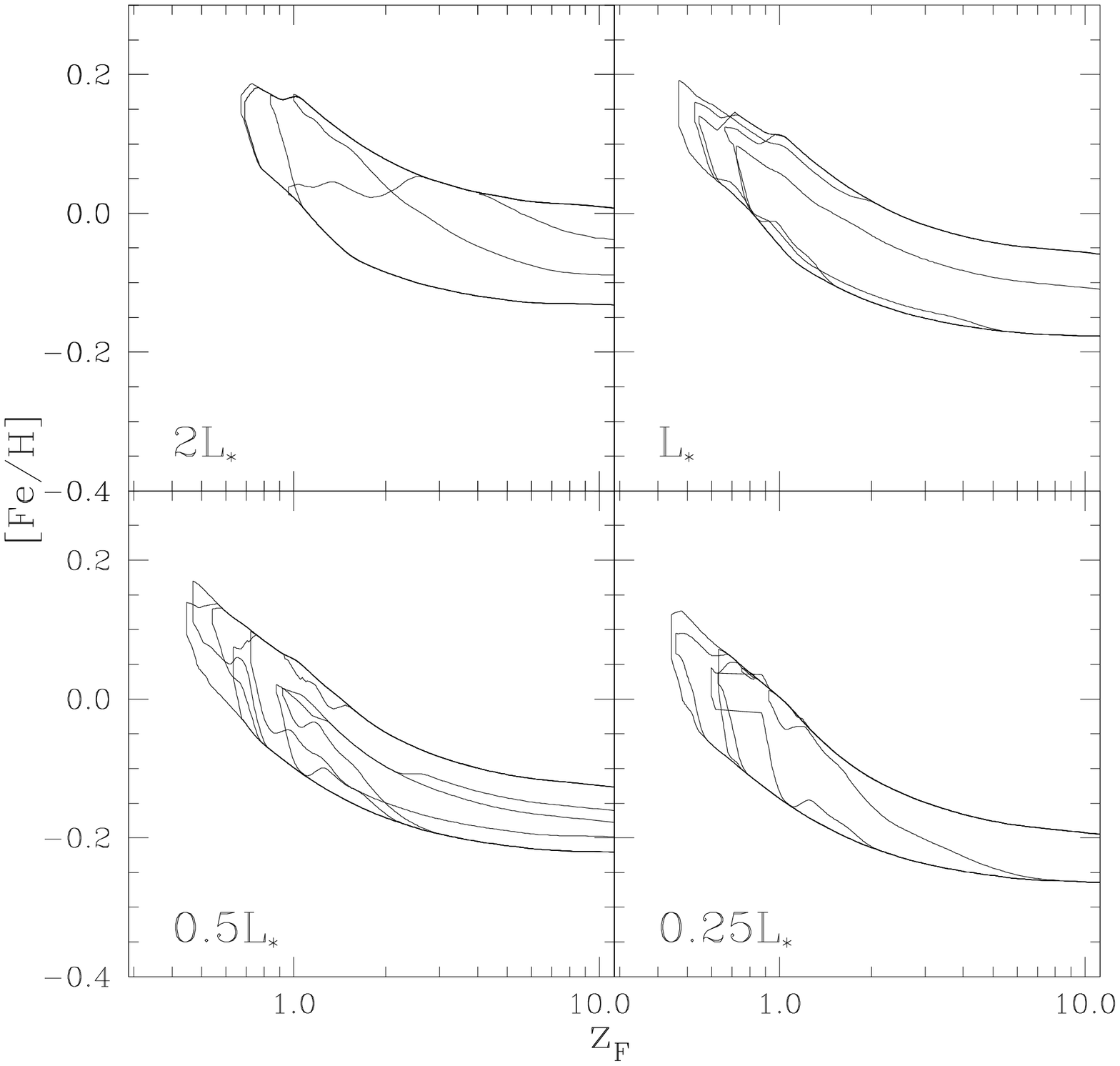}
\figcaption[fig3a.eps]
{{\bf a.} Constraints on the iron abundance [Fe/H] and formation redshift
$z_F$ of early-type galaxies in clusters in the redshift range $0.3<z<0.5$
(see text). The iron abundance is computed as [Fe/H]$=\log(Z/X)-\log(Z_\odot
/X_\odot)$, with $Z_\odot=0.02$. The four panels correspond to four luminosity
bins. Different contours show the areas allowed for different clusters
(Table~2).
  \label{f3a}}
\setcounter{figure}{2}

\clearpage
\plotone{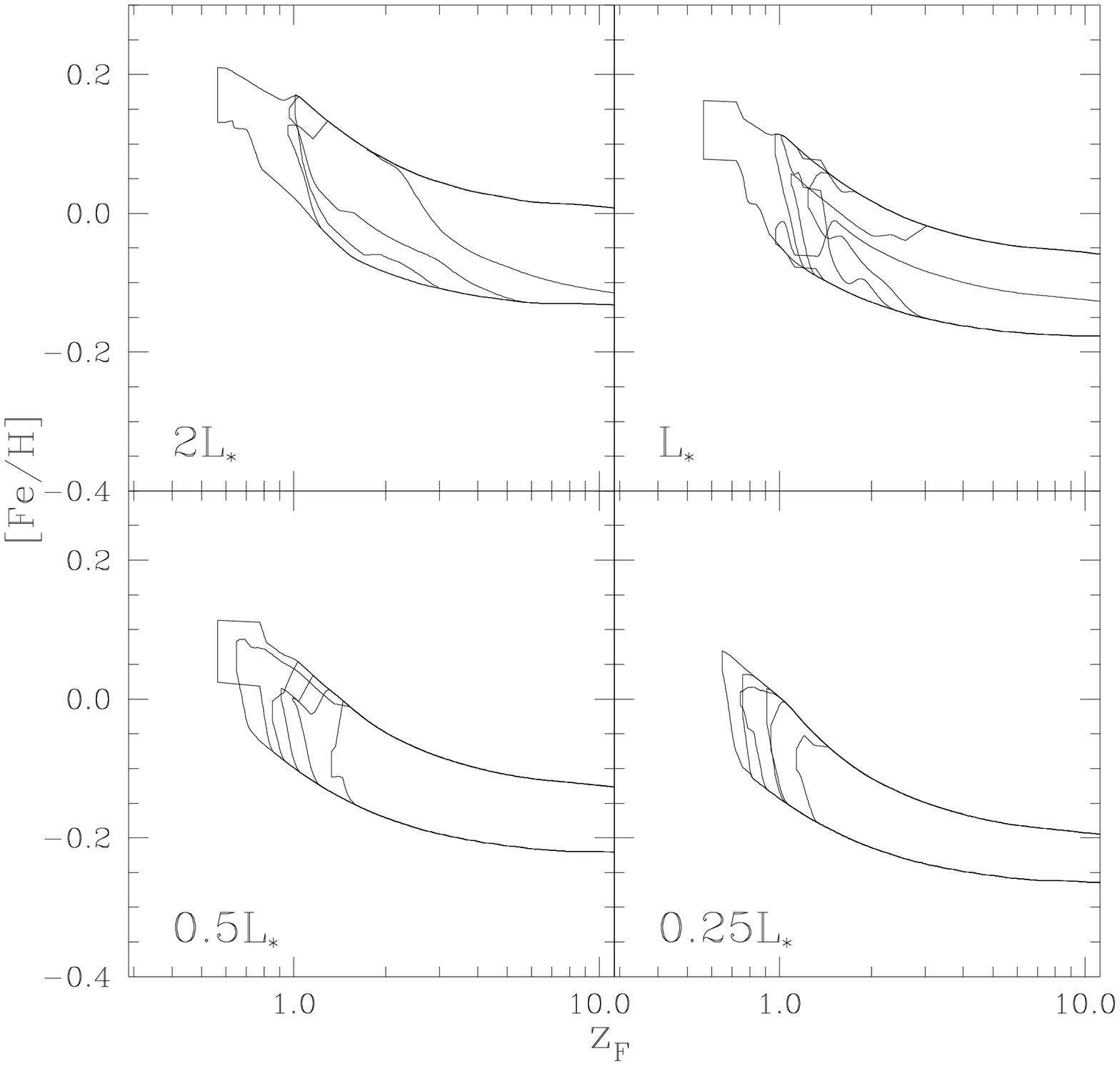}
\figcaption[fig3b.eps]
{{\bf b.} Same as Fig.~3{\it a} but for clusters in the redshift range
$0.5<z<0.7$ (Table~2).
   \label{f3b}}
\setcounter{figure}{2}

\clearpage
\plotone{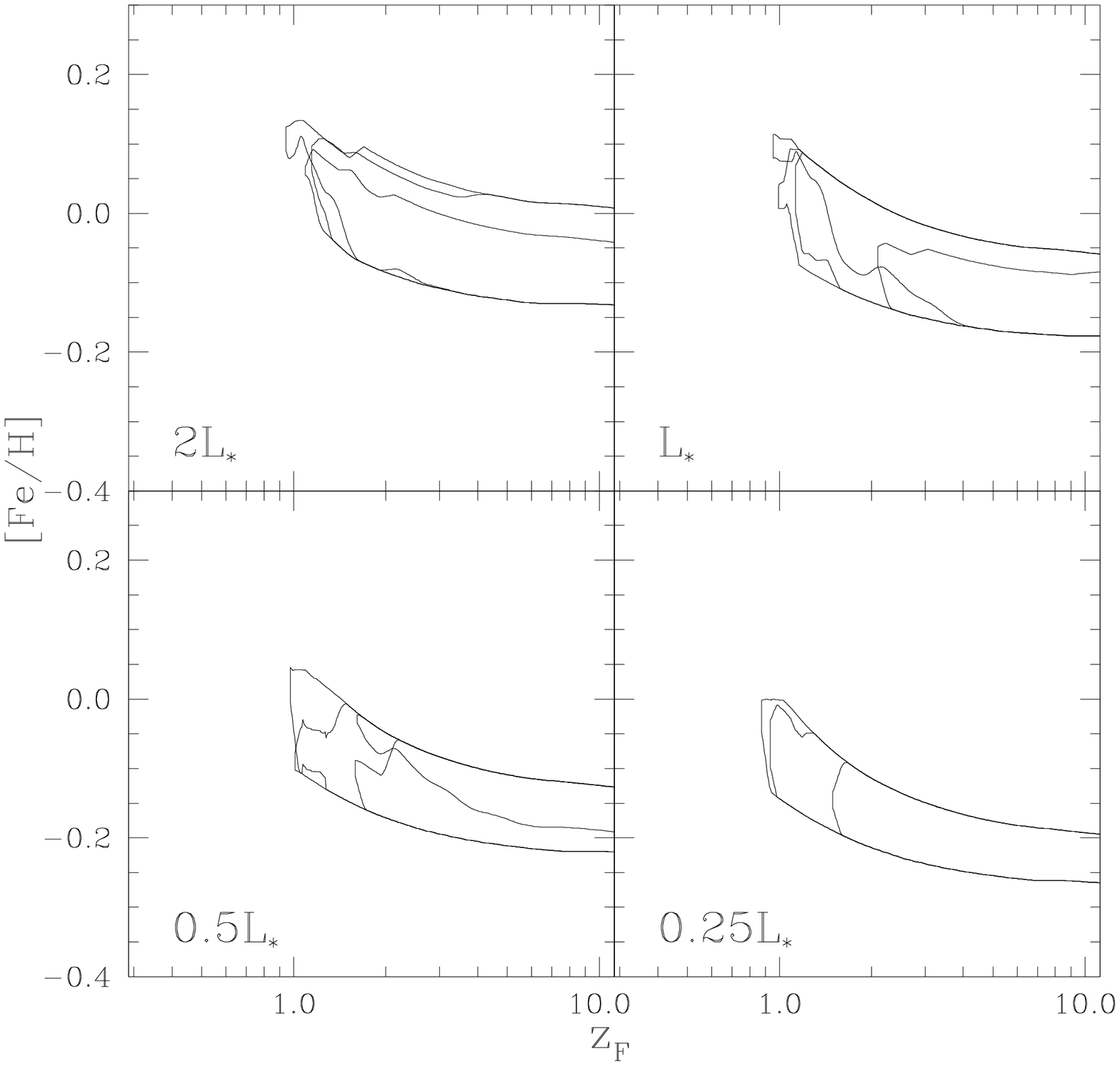}
\figcaption[fig3c.eps]
{{\bf c.} Same as Fig.~3{\it a} but for clusters in the redshift range
$0.7<z<1.0$ (Table~2).
   \label{f3c}}

\clearpage
\plotone{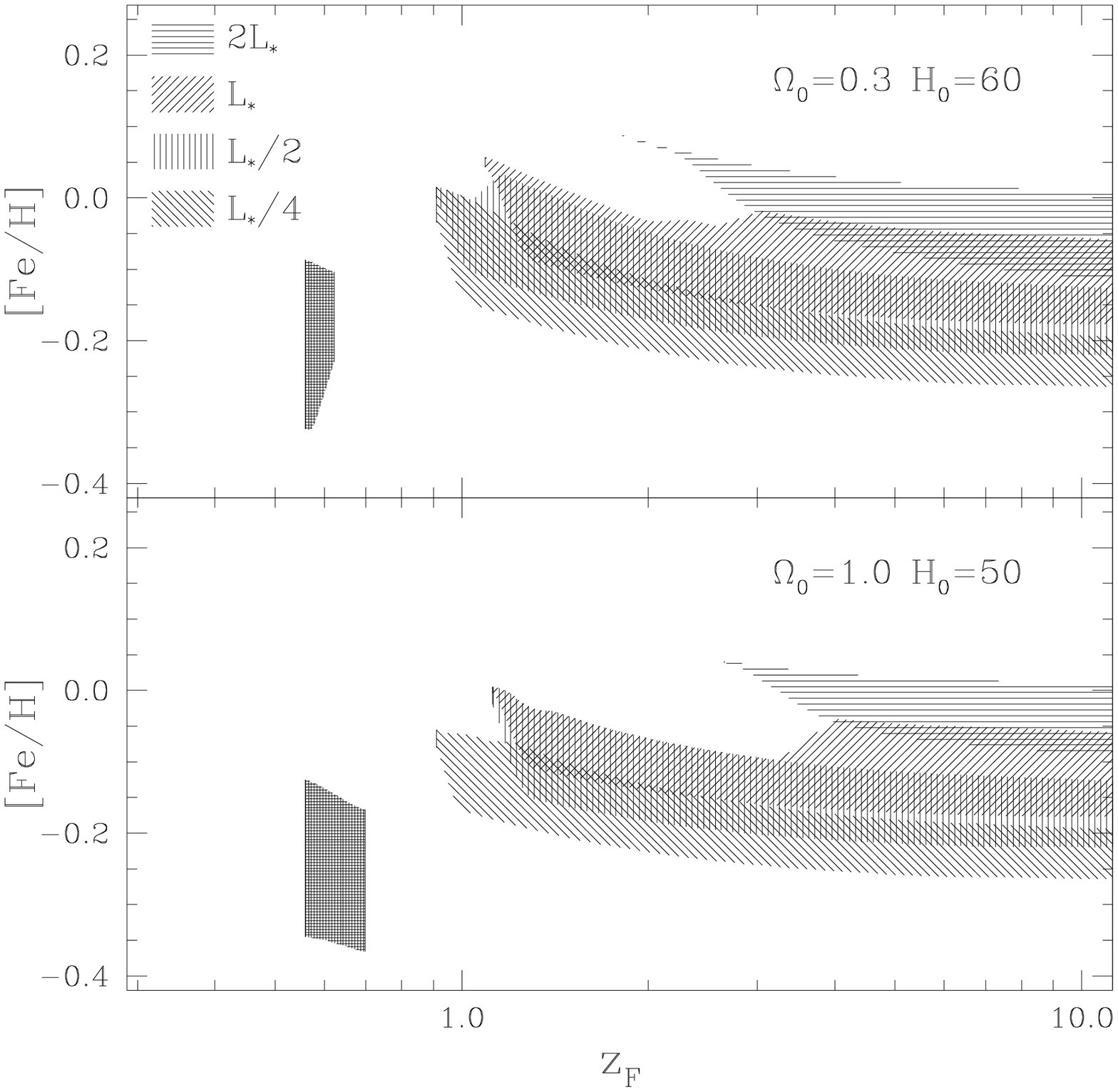}
\figcaption[fig4.eps]
{Constraints on the iron abundance [Fe/H] and formation redshift $z_F$ of
early-type galaxies in the cluster Cl~0016+16 at $z=0.545$. Four different
shadings refer to the four luminosity bins of Figure~3. The heavy
shaded region towards low $z_F$ maps the faint blue outliers on the assumption
that they are cluster members. The two panels show the results for two
different cosmological models, as indicated.
\label{f4}}

\clearpage
\plotone{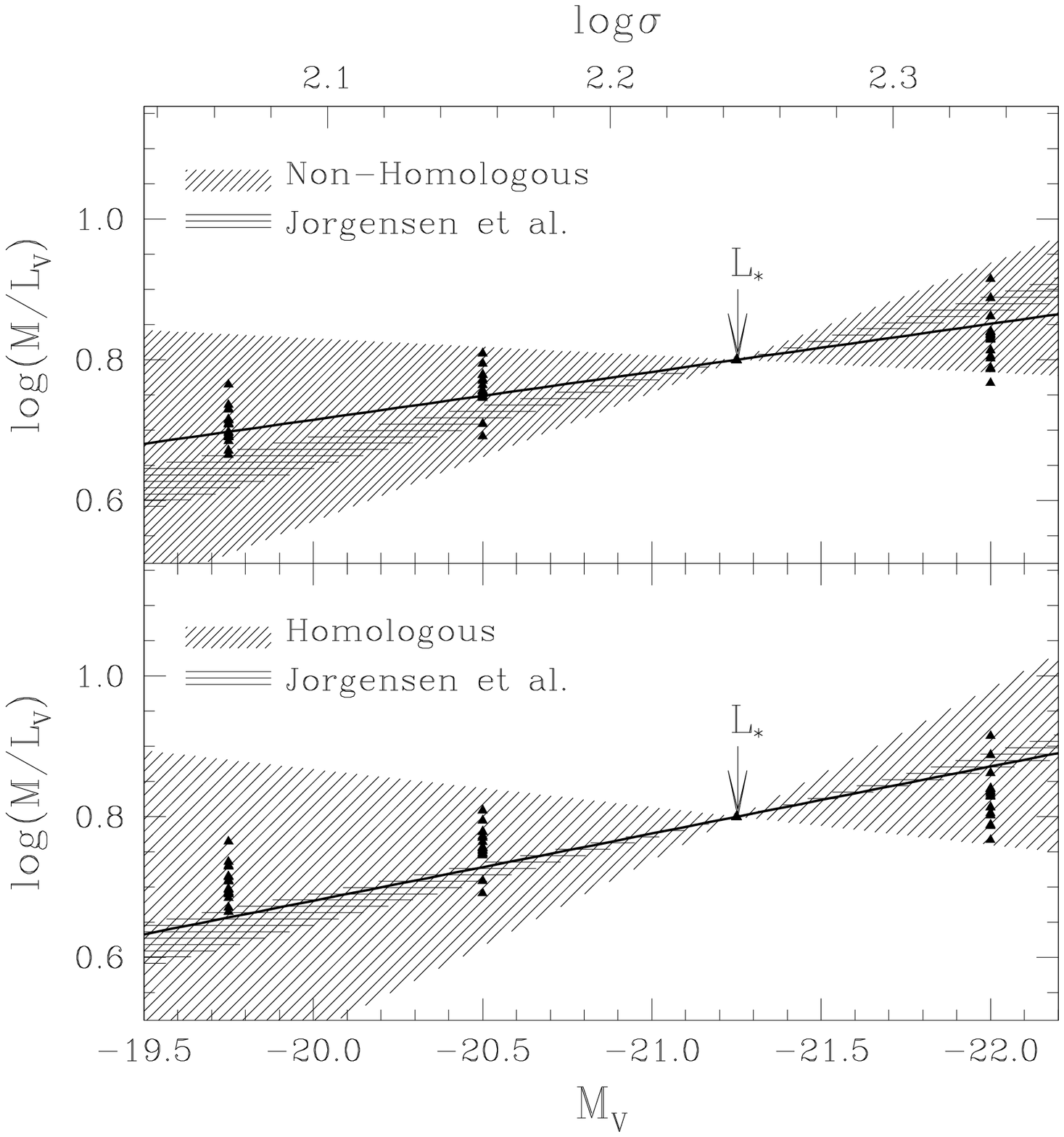}
\vskip-0.5truecm
\figcaption[fig5.eps]
{Mean mass-to-$V$ light ratios predicted for early-type galaxies of
different luminosities in the clusters of Table~2 ({\it filled
triangles}). The results are normalized to $\log(M/L_V)= 0.8$ at $L_V
=L_\ast$. The shaded regions indicate the dispersions around the relations
obtained by J\o rgensen et al. (1996) and Graham \& Colless (1997) from
independent observations; in the latter case, two fits are shown corresponding
to non-homologous ($\propto r^{1/n}$, {\it upper panel}~) and homologous
($\propto r^{1/4}$, {\it bottom panel}~) galaxy light profiles, the heavy
solid line indicating the mean relation (see text for details).
\label{f5}}

\clearpage
\plotone{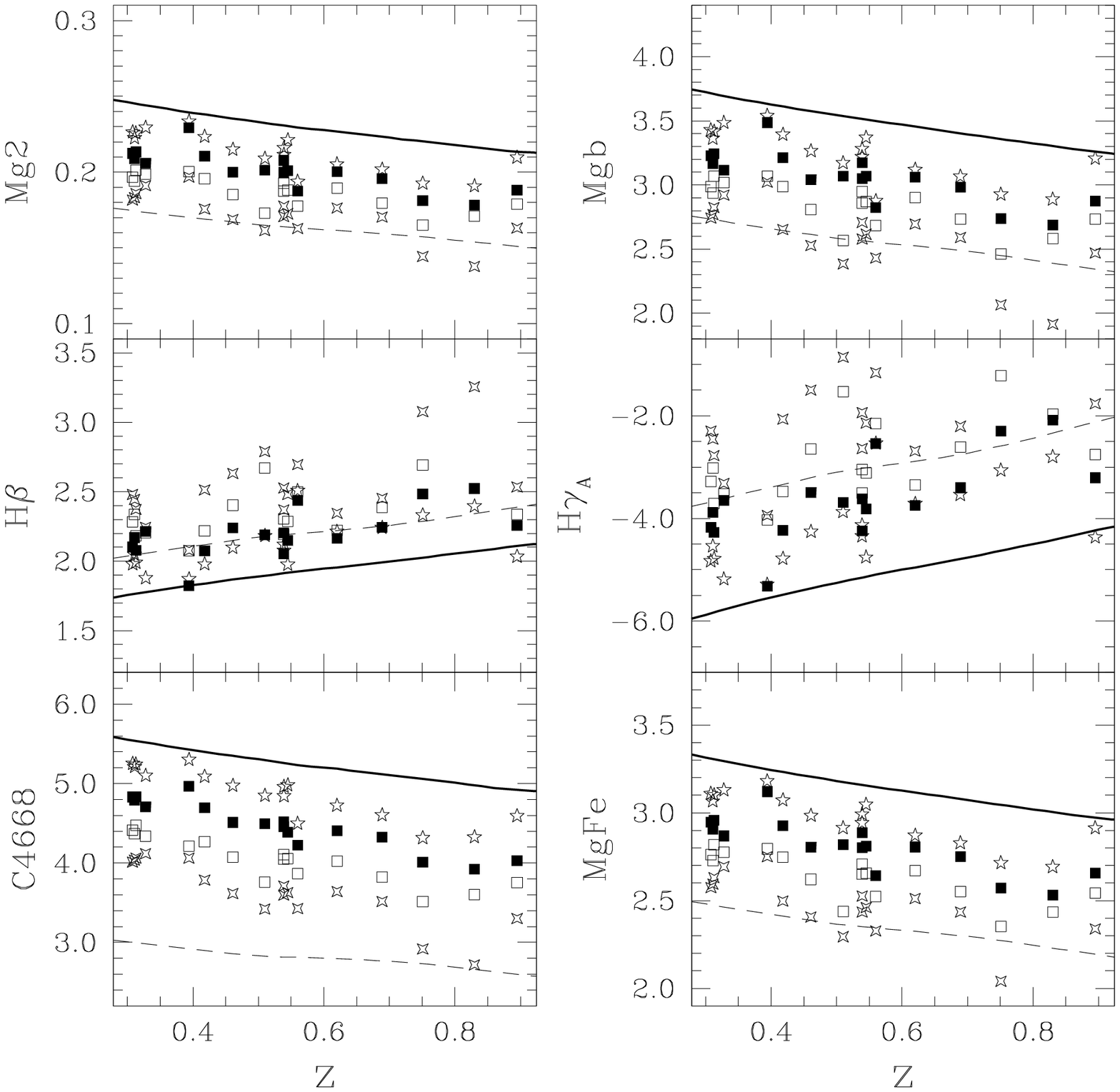}
\figcaption[fig6.eps]
{Mean Mg$_2$, H$\beta$, C4668, Mg$_b$, H$\gamma_A$ and MgFe indices
predicted for early-type galaxies in the clusters of Table~2. The four 
different symbols correspond to four luminosity bins ({\it 5-point
star}: $2L_\ast$; {\it filled square}: $L_\ast$; {\it hollow square}:
$0.5L_\ast$; {\it 4-point star}: $0.25L_\ast$). The lines show passive
evolution models with $z_F=5$ and the metallicities $Z_\odot$ ({\it
solid}) and $0.4Z_\odot$ ({\it dashed}).
\label{f6}}

\end{document}